\documentclass{article}
\usepackage{graphicx}
\usepackage{amsmath}

\input{tcilatex}

\begin{document}

\title{\textbf{2T Physics and Quantum Mechanics}}
\author{W. Chagas-Filho \\
Physics Department, Federal University of Sergipe, Brazil}
\maketitle

\begin{abstract}
We use a local scale invariance of a classical Hamiltonian and describe how
to construct six different formulations of quantum mechanics in spaces with
two time-like dimensions. All these six formulations have the same classical
limit described by the same Hamiltonian. One of these formulations is used
as a basis for a complementation of the usual quantum mechanics when in the
presence of gravity. 
\end{abstract}

\section{Introduction}

\noindent\ The wave-particle duality of matter and energy is one of the most
fundamental aspects of physics. The far reaching theoretical implications of
the existence of the wave-particle duality are not completely understood
until now. The best known implication of this duality is that quantum
mechanics can be equivalently formulated in the coordinate representation
and in the momentum representation. While the coordinate representation
emphasizes the particle aspect by assuming a defined position, the momentum
representation is related to the wave aspect because the magnitude $p$ of
the momentum of a particle is directly related to the wave length $\lambda $
of the associated wave by the de Broglie relation $p=\frac{h}{\lambda }$,
where $h$ is Planck%
\'{}%
s constant.

Some years ago it was discovered [1] that this complementarity of the
descriptions in terms of coordinates and momenta of quantum mechanics can be
made explicit as a classical local symmetry of an action functional
describing the motion of a massless scalar relativistic particle in a
space-time with an extra space-like dimension and an extra time-like
dimension. For the purpose of this paper, which is to further investigate
the theoretical implications of the complementarity of the wave and particle
aspects of matter and energy, the interesting aspect of this new physics
[2-18] with two time-like dimensions (2T physics) is that the duality of
coordinates and momenta appears already at\ the classical level, and this
makes it easier to follow its implications because the quantum mechanical
ordering ambiguities are absent. In this paper we use the local
indistinguishability of coordinate and momentum in 2T physics to suggest a
complementation of the basic equations of quantum mechanics.

In a previous paper [19], we presented a finite local scale invariance of
the 2T physics Hamiltonian and showed how this local invariance can be used
to relate the $d+2$ dimensional Minkowski space of 2T physics to a
Riemannian space of the same dimensionality. Although changing from a flat
space to a curved position dependent space using a local invariance is
already an interesting observation, it is not the only one. The finite local
scale invariance of the 2T Hamiltonian also associates to the $d+2$
dimensional Minkowski space of 2T physics another $d+2$ dimensional
Riemannian space where the geometry is described by a momentum dependent
tensor. More surprising is that the Hamiltonian equations of motion are
identical in these three spaces.

In the usual one-time (1T) physics, position dependent metric tensors play
an important role in the most general position space formulation of quantum
mechanics [20]. In this general formulation, these tensors appear in the
spectral decomposition of the unity, define the correct integration measure
for the inner product and are present in the most general expression of the
position matrix elements for self adjoint momentum operators in position
space [20] 
\begin{equation*}
\langle x\mid \hat{p}_{\alpha }\mid x%
{\acute{}}%
\rangle =\frac{i\hbar }{g^{\frac{1}{4}}(x)}\frac{\partial }{\partial
x^{\alpha }}[\frac{1}{g^{\frac{1}{4}}(x)}\delta ^{n}(x-x%
{\acute{}}%
)]
\end{equation*}
\begin{equation}
+\frac{1}{\sqrt{g(x)}}A_{\alpha }(x)\delta ^{n}(x-x%
{\acute{}}%
)  \tag{1.1}
\end{equation}
where $g(x)=\det g_{\alpha \beta }(x)$ and $\alpha ,\beta =1,...,n$. Since
quantum mechanics can be equivalently formulated in the position or in the
momentum representation, the appearance in 2T physics of a momentum
dependent metric tensor may be considered as an indication that the momentum
space versions of quantum mechanical equations such as (1.1) and others are
still lacking. The construction of these momentum space equations is one of
the motivations for this paper.

As can be seen in (1.1), the other central object in the general position
space formulation of quantum mechanics described in [20] is the vector field 
$A_{\alpha }(x)$. It has a vanishing strength tensor, 
\begin{equation}
F_{\alpha \beta }=\frac{\partial A_{\beta }}{\partial x^{\alpha }}-\frac{%
\partial A_{\alpha }}{\partial x^{\beta }}=0  \tag{1.2}
\end{equation}
and because of this condition it defines a section of a flat U(1) bundle
over the position space. The vector field is present only if the position
space has a non-trivial topology. In position spaces with trivial topology $%
A_{\alpha }(x)$ can always be gauged away [20].

In this paper we use the results of [19], together with new results that are
presented here, to show how one can extend to momentum space the general
position space formulation of quantum mechanics described in [20]. As we
will see here, in addition to the concept of a momentum dependent metric
tensor which was found necessary in [19], this paper evidentiates the need
for the concept of a momentum dependent vector field. Combining these two
concepts we can write down the basic equations for a formulation of quantum
mechanics in total agreement with the wave-particle duality. This paper also
gives relevant contributions to the development of 2T physics with vector
fields. In addition to the local scale invariance of the classical
Hamiltonian equations of motion in the case when $A_{M}=A_{M}(X)$ that was
presented in [19], this paper also points out that the classical Hamiltonian
equations of motion are invariant under local scale transformations in the
case when $A_{M}=A_{M}(P)$. For each of the symmetries of the 2T action in
the case when $A_{M}=A_{M}(X)$ that were presented in [19], this paper also
presents the corresponding symmetry with $A_{M}=A_{M}(P)$. This existence of
the same symmetries of the 2T action in position space and in momentum space
is what gives support for the new quantum mechanical equations we write here.

The paper is organized as follows. In section two we briefly review how we
can use a local scale invariance of the massless scalar relativistic
particle Hamiltonian to introduce a new bracket structure in phase space.
These new brackets are the classical analogues of the Snyder commutators
[21] in the case where the noncommutativity parameter is $\theta =1$. The
Snyder commutators were derived in 1947 in a projective geometry approach to
the de Sitter (dS) space in the momentum representation and are considered
here as the first evidence for momentum dependent metric tensors in quantum
mechanics in the presence of gravity.

The case of momentum dependent metric tensors was not included in the
general formulation of quantum mechanics presented in [20]. However, the
necessity for this kind of tensor field in quantum mechanics is clearly
suggested in the most general expression for the wave function $\langle
x\mid p\rangle $ obtained in [20]. In this paper we adopt the point of view
that momentum dependent tensor fields are necessary in quantum mechanics in
the presence of gravity as a consequence of the wave-particle duality and in
section three we briefly review how we can use the local
indistinguishability of position and momentum in 2T physics to show that
momentum dependent metric tensors also have a natural existence in $d+2$
dimensions.

Section four presents the basic equations of a formulation of quantum
mechanics that completely incorporates the wave-particle duality. This is
done by introducing the corresponding momentum space expressions of the
position space expressions obtained in [20]. A difficulty that appears in
this formulation of quantum mechanics is that we also need the concept of a
momentum dependent vector field. As a basis for this concept, in section
five we present the action that describes in a unified way 2T physics with
position dependent or momentum dependent vector fields. For each symmetry of
our action in position space we present the corresponding symmetry in
momentum space. We also discuss the scale invariance of the equations of
motion when $A_{M}=A_{M}(X)$ and when $A_{M}=A_{M}(P)$ and conclude that
there are six possible equivalent formulations of quantum mechanics that
have the same classical Hamiltonian limit described by 2T physics. Further
concluding remarks appear in section six.

\section{Massless Relativistic Particles}

In this section we briefly review how the classical analogues of the Snyder
commutators, obtained in 1947 for the dS space in the momentum
representation, can be derived in massless scalar relativistic particle
theory using a local scale invariance of the Hamiltonian. For details the
reader should see [19].

A massless scalar relativistic particle in a $d$-dimensional Minkowski
space-time with signature $(d-1,1)$ is described by the action 
\begin{equation}
S=\frac{1}{2}\int_{\tau _{i}}^{\tau _{f}}d\tau \lambda ^{-1}\dot{x}^{2} 
\tag{2.1}
\end{equation}
where $\lambda (\tau )$ is an auxiliary variable, $x^{\mu }=x^{\mu }(\tau )$%
, $\dot{x}^{2}=\dot{x}^{\mu }\dot{x}^{\nu }\eta _{\mu \nu }$ and $\eta _{\mu
\nu }$ is the flat Minkowski metric. A dot denotes derivatives with respect
to the parameter $\tau $. Action (2.1) is invariant under the local
infinitesimal reparameterizations 
\begin{equation*}
\delta x_{\mu }=\alpha (\tau )\dot{x}_{\mu }\text{ \ \ \ \ \ }\delta \lambda
=\frac{d}{d\tau }[\alpha (\tau )\lambda ]
\end{equation*}
and therefore describes gravity on the world-line. In the transition to the
Hamiltonian formalism action (2.1) gives the canonical momenta 
\begin{equation}
p_{\lambda }=0  \tag{2.2}
\end{equation}
\begin{equation}
p_{\mu }=\frac{\dot{x}_{\mu }}{\lambda }  \tag{2.3}
\end{equation}
and the canonical Hamiltonian 
\begin{equation}
H=\frac{1}{2}\lambda p^{2}  \tag{2.4}
\end{equation}
Equation (2.2) is a primary constraint [22]. Introducing the Lagrange
multiplier $\xi (\tau )$ for this constraint we can write the Dirac
Hamiltonian 
\begin{equation}
H_{D}=\frac{1}{2}\lambda p^{2}+\xi p_{\lambda }  \tag{2.5}
\end{equation}
Requiring the dynamical stability of constraint (2.2), $\dot{p}_{\lambda
}=\{p_{\lambda },H_{D}\}=0$, and using the Poisson bracket $\{\lambda
,p_{\lambda }\}=1$, we obtain the secondary constraint 
\begin{equation}
\phi =\frac{1}{2}p^{2}\approx 0  \tag{2.6}
\end{equation}
Constraints (2.2) and (2.6) have vanishing Poisson bracket and are therefore
first class constraints [22]. Constraint (2.2) generates translations in the
arbitrary variable $\lambda (\tau )$ and can be dropped from the formalism.
The notation $\approx $ means that $\phi $ \textsl{weakly vanishes} [23].
Weak equalities hold over the entire phase space and can be turned into
strong equalities on the constraint surface.

Action (2.1) can be rewritten in Hamiltonian form as 
\begin{equation}
S=\int_{\tau _{i}}^{\tau _{f}}d\tau (\dot{x}.p-\frac{1}{2}\lambda p^{2}) 
\tag{2.7}
\end{equation}
The first class constraint (2.6) generates the gauge transformations 
\begin{equation}
\delta x_{\mu }=\epsilon (\tau )\{x_{\mu },\phi \}=\epsilon (\tau )p_{\mu } 
\tag{2.8a}
\end{equation}
\begin{equation}
\delta p_{\mu }=\epsilon (\tau )\{p_{\mu },\phi \}=0  \tag{2.8b}
\end{equation}
\begin{equation}
\delta \lambda =\dot{\epsilon}(\tau )  \tag{2.8c}
\end{equation}
computed in terms of the Poisson brackets 
\begin{equation}
\{x_{\mu },x_{\nu }\}=0\text{ \ \ \ }\{p_{\mu },p_{\nu }\}=0\text{ \ \ \ }%
\{x_{\mu },p_{\nu }\}=\eta _{\mu \nu }  \tag{2.9}
\end{equation}
under which action (2.7) transforms as 
\begin{equation}
\delta S=\int_{\tau _{i}}^{\tau _{f}}d\tau \frac{d}{d\tau }(\epsilon \phi ) 
\tag{2.10}
\end{equation}
Since the interval $(\tau _{i},\tau _{f})$ is arbitrary, action (2.7) is
invariant under transformation (2.8) and the quantity $Q=\epsilon \phi $ can
be interpreted as the conserved Hamiltonian Noether charge or as the
generator of the gauge transformations (2.8), depending on wether the
equations of motion are satisfied or not [24]. This property of Hamiltonian
Noether charges $Q$ will be used to confirm some of the results contained in
this paper.

The gravitational field, regarded as a gauge field [25], can correspond to
several symmetry groups: 1) the general covariant group, 2) the local
Lorentz group, and 3) the group of scale transformations of the interval. In
the first case the properties of the gravitational field are determined by
the properties of the metric tensor, and this gives the usual Einstein
theory. In the second case they are determined by the properties of the
Ricci connection coefficients and this leads to equations of the fourth
order. In the third case, it is assumed that the source of the gravitational
field is the trace of the energy-momentum tensor and that the carries are
scalar particles [25]. In agreement with the third point of view, the
massless particle Hamiltonian (2.4) is invariant under the finite local
scale transformations 
\begin{equation}
p_{\mu }\rightarrow \exp \{-\beta (\tau )\}p_{\mu }  \tag{2.11a}
\end{equation}
\begin{equation}
\lambda \rightarrow \exp \{2\beta (\tau )\}\lambda  \tag{2.11b}
\end{equation}
where $\beta (\tau )$ is an arbitrary scalar function. From equation (2.3)
for the canonical momentum we find that $x^{\mu }$ transforms as 
\begin{equation}
x_{\mu }\rightarrow \exp \{\beta (\tau )\}x_{\mu }  \tag{2.11c}
\end{equation}
when $p_{\mu }$ transforms as in (2.11a). We can use the arbitrary character
of the function $\beta $ in the local scale invariance (2.11), together with
the first class property of constraint (2.6), to change to a bracket
structure different from the usual Poisson brackets (2.9). The simplest
possibility is to choose $\beta =\frac{1}{2}p^{2}$. In this gauge the phase
space has the bracket structure [19] 
\begin{equation}
\{p_{\mu },p_{\nu }\}\approx 0  \tag{2.12a}
\end{equation}
\begin{equation}
\{x_{\mu },p_{\nu }\}\approx \eta _{\mu \nu }-p_{\mu }p_{\nu }  \tag{2.12b}
\end{equation}
\begin{equation}
\{x_{\mu },x_{\nu }\}\approx -M_{\mu \nu }  \tag{2.12c}
\end{equation}
where $M_{\mu \nu }=x_{\mu }p_{\nu }-x_{\nu }p_{\mu }$ is the generator of
Lorentz transformations. It can be verified that all Jacobi identities among
the canonical variables still close if we use the brackets (2.12) instead of
the Poisson brackets (2.9). In the transition to the quantized theory using
the correspondence principle rule that [commutator]=ih\{bracket\}, the
brackets (2.12) will reproduce the Snyder commutators [21] for the dS space
in the momentum representation and with noncommutativity parameter $\theta
=1 $.

In the presence of gravity and at length scales near the Planck length, the
fundamental commutator $[x_{\mu },p_{\nu }]=ih\eta _{\mu \nu }$ of quantum
mechanics must be replaced [26] by a more general commutator $[x_{\mu
},p_{\nu }]=ihg_{\mu \nu }$. This is because the large amounts of
relativistic momentum involved in the quantum measurement processes
necessarily modify the space-time geometry at these length scales [26]. $%
g_{\mu \nu }$ is in principle a function of the positions, but this paper
calls attention to the fact that $g_{\mu \nu }$ can also be a momentum
dependent function. This can be seen in the bracket (2.12b) we derived for
the massless relativistic particle. This can also be seen in the old Snyder
commutators for the dS space in the momentum representation.

Hamiltonian (2.4) generates the classical equations of motion 
\begin{equation}
\dot{x}_{\mu }=\{x_{\mu },H\}=\lambda p_{\mu }  \tag{2.13a}
\end{equation}
\begin{equation}
\dot{p}_{\mu }=\{p_{\mu },H\}=0  \tag{2.13b}
\end{equation}
computed in terms of the Poisson brackets (2.9). If we now change to the
momentum dependent background 
\begin{equation}
\bar{g}_{\mu \nu }=\eta _{\mu \nu }-p_{\mu }p_{\nu }  \tag{2.14}
\end{equation}
implied by bracket (2.12b), the new Hamiltonian $\bar{H}$ in this background
is given by 
\begin{equation}
\bar{H}=H-2\lambda \phi ^{2}  \tag{2.15}
\end{equation}
The Hamiltonian (2.15) in the background (2.14) differs from (2.4) by a term
that is quadratic in constraint (2.6). This term can be dropped and the new
Hamiltonian in the background (2.14) is identical to (2.4) in Minkowski
space.

Although the Hamiltonians are identical, in the background (2.14) the
Poisson brackets (2.9) are no longer valid. They must be replaced by
brackets (2.12). Computing the equations of motion using the Hamiltonian
(2.4) and brackets (2.12) we find 
\begin{equation}
\dot{x}_{\mu }=\{x_{\mu },H\}=\lambda p_{\mu }-2\lambda p_{\mu }\phi 
\tag{2.16a}
\end{equation}
\begin{equation}
\dot{p}_{\mu }=\{p_{\mu },H\}=0  \tag{2.16b}
\end{equation}
We see that the equations of motion in the background (2.14) differ from
(2.13) by a term that is linear in the constraint $\phi $. Again, this term
can be dropped. From these observations we may conclude that, at the
classical level, the massless particle Hamiltonian dynamics in the momentum
space background (2.14) is indistinguishable from the Hamiltonian dynamics
in Minkowski space.

\section{2T Physics}

In this section we consider how the position space and momentum space
higher-dimensional extensions of brackets (2.12) can be obtained from a
local scale invariance of the 2T Hamiltonian. We also consider the
invariance f the classical equations of motion in the corresponding
backgrounds.

The construction of 2T physics [1-18] is based on the introduction of a new
gauge invariance in phase space by gauging the duality of the quantum
commutator $[X_{M},P_{N}]=ih\eta _{MN}$. This procedure leads to a
symplectic $Sp(2,R)$ gauge theory. To remove the distinction between
position and momentum we rename them $X_{1}^{M}=X^{M}(\tau )$ and $%
X_{2}^{M}=P^{M}(\tau )$ and define the doublet $X_{i}^{M}(\tau
)=(X_{1}^{M},X_{2}^{M})$. The local $Sp(2,R)$ symmetry acts as 
\begin{equation}
\delta X_{i}^{M}(\tau )=\epsilon _{ik}\omega ^{kl}(\tau )X_{l}^{M}(\tau ) 
\tag{3.1}
\end{equation}
$\omega ^{ij}(\tau )$ is a symmetric matrix containing three local
parameters and $\epsilon _{ij}$ is the Levi-Civita symbol that serves to
raise or lower indices. The $Sp(2,R)$ gauge field $A^{ij}$ is symmetric in $%
(i,j)$ and transforms as 
\begin{equation}
\delta A^{ij}=\partial _{\tau }\omega ^{ij}+\omega ^{ik}\epsilon
_{kl}A^{lj}+\omega ^{jk}\epsilon _{kl}A^{il}  \tag{3.2}
\end{equation}
The covariant derivative is 
\begin{equation}
D_{\tau }X_{i}^{M}=\partial _{\tau }X_{i}^{M}-\epsilon _{ik}A^{kl}X_{l}^{M} 
\tag{3.3}
\end{equation}
An action invariant under the $Sp(2,R)$ gauge symmetry is 
\begin{equation}
S=\frac{1}{2}\int d\tau (D_{\tau }X_{i}^{M})\epsilon ^{ij}X_{j}^{N}\eta _{MN}
\tag{3.4a}
\end{equation}
After an integration by parts this action can be written as 
\begin{equation*}
S=\int d\tau (\partial _{\tau }X_{1}^{M}X_{2}^{N}-\frac{1}{2}%
A^{ij}X_{i}^{M}X_{j}^{N})\eta _{MN}
\end{equation*}
\begin{equation}
=\int d\tau \lbrack \dot{X}.P-(\frac{1}{2}\lambda _{1}P^{2}+\lambda _{2}X.P+%
\frac{1}{2}\lambda _{3}X^{2})]  \tag{3.4b}
\end{equation}
where $A^{11}=\lambda _{3}$, $A^{12}=A^{21}=\lambda _{2}$, \ $A^{22}=\lambda
_{1}$ and the canonical Hamiltonian is 
\begin{equation}
H=\frac{1}{2}\lambda _{1}P^{2}+\lambda _{2}X.P+\frac{1}{2}\lambda _{3}X^{2} 
\tag{3.5}
\end{equation}
The equations of motion for the $\lambda $'s give the first class
constraints 
\begin{equation}
\phi _{1}=\frac{1}{2}P^{2}\approx 0  \tag{3.6}
\end{equation}
\begin{equation}
\phi _{2}=X.P\approx 0  \tag{3.7}
\end{equation}
\begin{equation}
\phi _{3}=\frac{1}{2}X^{2}\approx 0  \tag{3.8}
\end{equation}
Constraints (3.6)-(3.8), as well as evidences of 2T physics, were
independently obtained in [27]. Equations (3.6) and (3.8) can be interpreted
as constraints only if the hypersurfaces $P_{0}=P_{1}=...=P_{d+1}=0$ and $%
X_{0}=X_{1}=...=X_{d+1}=0$ are removed from phase space [27]. Only in this
case we have a consistent Hamiltonian formalism defined over a regular [23]
constraint surface. This removal of the origin of phase space induces a
non-trivial phase space topology. In the case of the usual position space of
1T physics, a non-trivial topology is associated with the presence of a
position dependent vector field in the quantized theory [20]. This vector
field has a vanishing strength tensor and defines a section of a flat U(1)
bundle over the position space [20]. The presence of general vector fields
in 2T physics will be considered in section five. The case of position
dependent vector fields with non-vanishing strength tensors was first
discussed in [5].

If we consider the usual Minkowski metric as the background space, we find
that the surface defined by the constraint equations (3.6)-(3.8) is trivial.
The metrics giving a non-trivial constraint surface, preserving the
unitarity of the theory, and avoiding the ghost problem are the metrics with
two time-like dimensions [1-18]. For the purposes of this paper it is best
start working in a Minkowski space with signature $(d,2).$ Action (3.4b) is
the $(d+2)$-dimensional extension of the $d$-dimensional massless particle
action (2.7). Action (3.4b) describes conformal gravity on the world-line
[28,29,1].

We now introduce the Poisson brackets 
\begin{equation}
\{P_{M},P_{N}\}=0\text{ \ \ \ \ }\{X_{M},X_{N}\}=0\text{ \ \ \ \ }%
\{X_{M},P_{N}\}=\eta _{MN}  \tag{3.9}
\end{equation}
It can then be checked that action (3.4b) is invariant under Lorentz $%
SO(d,2) $ transformations with generator $L_{MN}=X_{M}P_{N}-X_{N}P_{M}$ 
\begin{equation}
\delta X_{M}=\frac{1}{2}\omega _{RS}\{L_{RS},X_{M}\}=\omega _{MR}X_{R} 
\tag{3.10a}
\end{equation}
\begin{equation}
\delta P_{M}=\frac{1}{2}\omega _{RS}\{L_{RS},P_{M}\}=\omega _{MR}P_{R} 
\tag{3.10b}
\end{equation}
\begin{equation}
\delta \lambda _{\alpha }=0,\text{ \ \ \ \ \ \ }\alpha =1,2,3  \tag{3.10c}
\end{equation}
under which $\delta S=0$. The $L_{MN}$ are gauge invariant because they have
vanishing brackets with constraints (3.6)-(3.8).

The first class constraints (3.6)-(3.8) generate the local transformations 
\begin{equation}
\delta X_{M}=\epsilon _{\alpha }(\tau )\{X_{M},\phi _{\alpha }\}=\epsilon
_{1}P_{M}+\epsilon _{2}X_{M}  \tag{3.11a}
\end{equation}
\begin{equation}
\delta P_{M}=\epsilon _{\alpha }(\tau )\{P_{M},\phi _{\alpha }\}=-\epsilon
_{2}P_{M}-\epsilon _{3}X_{M}  \tag{3.11b}
\end{equation}
\begin{equation}
\delta \lambda _{1}=\dot{\epsilon}_{1}+2\epsilon _{2}\lambda _{1}-2\epsilon
_{1}\lambda _{2}  \tag{3.11c}
\end{equation}
\begin{equation}
\delta \lambda _{2}=\dot{\epsilon}_{2}+\epsilon _{3}\lambda _{1}-\epsilon
_{1}\lambda _{3}  \tag{3.11d}
\end{equation}
\begin{equation}
\delta \lambda _{3}=\dot{\epsilon}_{3}+2\epsilon _{3}\lambda _{2}-2\epsilon
_{2}\lambda _{3}  \tag{3.11e}
\end{equation}
under which 
\begin{equation}
\delta S=\int_{\tau _{i}}^{\tau _{f}}d\tau \frac{d}{d\tau }(\epsilon
_{\alpha }\phi _{\alpha })  \tag{3.12}
\end{equation}
Similarly to the massless particle case, since the interval $(\tau _{i},\tau
_{f})$ is arbitrary, the quantity $Q=\epsilon _{\alpha }\phi _{\alpha }$,
with $\alpha =1,2,3$, can be interpreted as the conserved Hamiltonian
Noether charge or as the generator of the local transformations (3.11),
depending on wether the equations of motion are satisfied or not [24].

The 2T Hamiltonian (3.5) is invariant under the finite local scale
transformations 
\begin{equation}
X_{M}\rightarrow \exp \{\beta (\tau )\}X_{M}  \tag{3.13a}
\end{equation}
\begin{equation}
P_{M}\rightarrow \exp \{-\beta (\tau )\}P_{M}  \tag{3.13b}
\end{equation}
\begin{equation}
\lambda _{1}\rightarrow \exp \{2\beta (\tau )\}\lambda _{1}  \tag{3.13c}
\end{equation}
\begin{equation}
\lambda _{2}\rightarrow \lambda _{2}  \tag{3.13d}
\end{equation}
\begin{equation}
\lambda _{3}\rightarrow \exp \{-2\beta (\tau )\}\lambda _{3}  \tag{3.13e}
\end{equation}
where $\beta (\tau )$ is an arbitrary scalar function. Now we can use the
scale transformation (3.13), together with the first class property of
constraints (3.6)-(3.8), and by choosing the arbitrary function to be $\beta
=\frac{1}{2}P^{2}$, arrive at the brackets [19] 
\begin{equation}
\{P_{M},P_{N}\}\approx 0  \tag{3.14a}
\end{equation}
\begin{equation}
\{X_{M},P_{N}\}\approx \eta _{MN}-P_{M}P_{N}  \tag{3.14b}
\end{equation}
\begin{equation}
\{X_{M},X_{N}\}\approx -L_{MN}  \tag{3.14c}
\end{equation}
Brackets (3.14) are the $(d+2)$-dimensional extensions of the $d$%
-dimensional momentum space brackets (2.12) we found for the massless
particle in the previous section. By choosing the arbitrary scalar function $%
\beta $ to be a function of $P_{M}(\tau )$, we arrived at the momentum space
brackets (3.14). But in 2T physics momentum and position are
indistinguishable variables. So, in 2T physics, there must exist a position
space version of brackets (3.14). This position space version can be reached
by choosing $\beta =\frac{1}{2}X^{2}$ in transformation (3.13). Using again
the first class property of constraints (3.6)-(3.8), we arrive at the
brackets [19] 
\begin{equation}
\{P_{M},P_{N}\}\approx L_{MN}  \tag{3.15a}
\end{equation}
\begin{equation}
\{X_{M},P_{N}\}\approx \eta _{MN}+X_{M}X_{N}  \tag{3.15b}
\end{equation}
\begin{equation}
\{X_{M},X_{N}\}\approx 0  \tag{3.15c}
\end{equation}
Notice that brackets (3.15) can not be obtained from brackets (3.14) by
performing the duality transformation $X_{M}\rightarrow P_{M}$, $%
P_{M}\rightarrow -X_{M}$ which leaves the quantum commutator $%
[X_{M},P_{N}]=ih\eta _{MN}$ invariant. This duality does not allow us to
perform a transition from brackets (3.14) to brackets (3.15). The transition
from (3.14) to (3.15) involves the local scale invariance (3.13) of the 2T
Hamiltonian (3.5). Therefore this transition involves conformal gravity on
the world-line.

If we choose $\beta (\tau )=0$ in transformation (3.13) we obtain the
Poisson brackets (3.9). Working with the Poisson brackets (3.9), or with
brackets (3.14), or with brackets (3.15) is a matter of gauge choice and
therefore these three sets of brackets must lead to equivalent results at
the classical level. This can be easily verified to be true. Using the
Poisson brackets (3.9) we find that the 2T Hamiltonian (3.5) generates the
classical equations of motion 
\begin{equation}
\dot{X}_{M}=\{X_{M},H\}=\lambda _{1}P_{M}+\lambda _{2}X_{M}  \tag{3.16a}
\end{equation}
\begin{equation}
\dot{P}_{M}=\{P_{M},H\}=-\lambda _{2}P_{M}-\lambda _{3}X_{M}  \tag{3.16b}
\end{equation}
After dropping terms proportional to the first class constraints
(3.6)-(3.8), we find that the 2T Hamiltonian (3.5) and the equations of
motion (3.16) remain invariant if we change to the background 
\begin{equation}
\bar{G}_{MN}=\eta _{MN}-P_{M}P_{N}  \tag{3.17}
\end{equation}
and simultaneously replace the Poisson brackets (3.9) by brackets (3.14).
The 2T Hamiltonian (3.5) and the equations of motion (3.16) also remain
invariant if we change to the background 
\begin{equation}
G_{MN}=\eta _{MN}+X_{M}X_{N}  \tag{3.18}
\end{equation}
and simultaneously replace the Poisson brackets (3.9) by brackets (3.15). As
a consequence of the local scale invariance (3.13) there are three
equivalent classical Hamiltonian formulations of 2T physics. The first
formulation is in the usual Minkowisk space of 2T physics using the standard
Poisson brackets (3.9). The second formulation is in the momentum space
background (3.17) using brackets (3.14). The third formulation is in the
position space background (3.18) using brackets (3.15).

\section{Extended Quantum Mechanics}

The results of the previous sections bring with them the possibility of a
deeper insight into the formal structure of quantum mechanics. The idea is
to explicitly incorporate into quantum mechanics the indistinguishability of
position and momentum of 2T physics. This can be done without the need of
extra dimensions. To extend quantum mechanics we must introduce an
additional assumption between assumptions A1 and A2 of reference [20]. The
formulation of quantum mechanics we present in this section is based in
three assumptions, of which the first and the third ones are identical to A1
and A2 in [20]. Our assumptions are

1) There exists a basis $\mid x\rangle $ of the position space which is
spanned by the eigenvalues of the position operators $\hat{x}^{\alpha }$ $%
(\alpha =1,2,...,n),$ whose domain of eigenvalues coincides with all the
possible values of the coordinates $x^{\alpha }$ parameterizing the position
space $M(x)$, 
\begin{equation*}
\hat{x}^{\alpha }\mid x\rangle =x^{\alpha }\mid x\rangle \text{ \ \ , \ \ }%
\{x^{\alpha }\}\in M(x)
\end{equation*}

2) There exists a basis $\mid p\rangle $ of the momentum space which is
spanned by the eigenvalues of the momentum operators $\hat{p}_{\alpha }$ $%
(\alpha =1,2,...,n),$ whose domain of eigenvalues coincides with all the
possible values of the momenta $p_{\alpha }$ parameterizing the momentum
space $D(p)$, 
\begin{equation*}
\hat{p}_{\alpha }\mid p\rangle =p_{\alpha }\mid p\rangle \text{ \ , \ }%
\{p_{\alpha }\}\in D(p)
\end{equation*}

3) The representation spaces of the algebra are endowed with a Hermitian
positive definite inner product $\langle .\mid .\rangle $ for which the
operators $\hat{x}^{\alpha }$ and $\hat{p}_{\alpha }$ are self-adjoint.

Now consider quantum mechanics in position space. The construction of
quantum mechanics describing the diffeomorphic-covariant representations of
the Heisenberg algebra in terms of topological classes of a flat U(1) bundle
over position space has the parameterization [20] of the inner product $%
\langle x\mid x^{\prime }\rangle $ 
\begin{equation}
\langle x\mid x^{\prime }\rangle =\frac{1}{\sqrt{g(x)}}\delta
^{n}(x-x^{\prime })  \tag{4.1}
\end{equation}
where $g(x)$ is an arbitrary positive definite function defined over the
position space $M$. For a Riemannian manifold the natural choice [20] for $g$
is the determinant of the metric tensor, $g=\det g_{\alpha \beta }(x)$.
Position dependent metric tensors naturally appear in this formulation of
quantum mechanics.

Equation (4.1) implies the spectral decomposition [20] of the identity
operator in the position eigenbasis $\mid x\rangle $ 
\begin{equation}
1=\int_{M}d^{n}x\sqrt{g(x)}\mid x\rangle \langle x\mid  \tag{4.2}
\end{equation}
which in turn leads to the position space wave function representations $%
\psi (x)=\langle x\mid \psi \rangle $ and $\langle \psi \mid x\rangle
=\langle x\mid \psi \rangle ^{\ast }=\psi ^{\ast }(x)$ of any state $\mid
\psi \rangle $ belonging to the Heisenberg algebra representation space, 
\begin{equation}
\mid \psi \rangle =\int_{M}d^{n}x\sqrt{g(x)}\psi (x)\mid x\rangle  \tag{4.3}
\end{equation}
\begin{equation}
\langle \psi \mid =\int_{M}d^{n}x\sqrt{g(x)}\psi ^{\ast }(x)\langle x\mid 
\tag{4.4}
\end{equation}
The inner product of two states $\mid \psi \rangle $ and $\mid \varphi
\rangle $ is then given in terms of their position space wave functions $%
\psi (x)$ and $\varphi (x)$ as 
\begin{equation}
\langle \psi \mid \varphi \rangle =\int_{M}d^{n}x\sqrt{g(x)}\psi ^{\ast
}(x)\varphi (x)  \tag{4.5}
\end{equation}
The most general position space wave function representations of the
position and momentum operators are [20] 
\begin{equation}
\langle x\mid \hat{x}_{\alpha }\mid \psi \rangle =x_{\alpha }\langle x\mid
\psi \rangle =x_{\alpha }\psi (x)  \tag{4.6a}
\end{equation}
\begin{equation}
\langle x\mid \hat{p}_{\alpha }\mid \psi \rangle =\frac{-i\hbar }{g^{1/4}(x)}%
\left[ \frac{\partial }{\partial x^{\alpha }}+\frac{i}{\hbar }A_{\alpha }(x)%
\right] g^{1/4}(x)\psi (x)  \tag{4.6b}
\end{equation}
The vector field $A_{\alpha }(x)$ is present only in the case of
topologically non-trivial position spaces [20]. It has a vanishing strength
tensor $F_{\alpha \beta }$ as given by (1.2) and is related to arbitrary
local phase transformations of the position eigenvectors 
\begin{equation}
\mid x^{\prime }\rangle =e^{\frac{i}{\hbar }\chi (x)}\mid x\rangle 
\tag{4.7a}
\end{equation}
when 
\begin{equation}
A_{\alpha }^{\prime }(x)=A_{\alpha }(x)+\frac{\partial \chi (x)}{\partial
x^{\alpha }}  \tag{4.7b}
\end{equation}
where $\chi (x)$ is an arbitrary scalar function. From the above equations
we see that in Riemannian spaces, where $g(x)=\det g_{\alpha \beta }$,
position dependent metric tensors play a central role in this formulation of
quantum mechanics. The other central object in this formulation is the
vector field $A_{\alpha }(x)$ of vanishing strength tensor.

Now we consider quantum mechanics in momentum space and extend the
construction in [20]. The normalization of the momentum eigenstates is
parameterized according to [20] 
\begin{equation}
\langle p\mid p^{\prime }\rangle =\frac{1}{\sqrt{h(p)}}\delta
^{n}(p-p^{\prime })  \tag{4.8}
\end{equation}
where $h(p)$ is an arbitrary positive definite function defined over the
domain $D(p)$ of the momentum eigenvalues. The authors in [20] do not go
beyond this point and do not consider the possible forms of the function $%
h(p)$. However, from our experience with the massless scalar relativistic
particle in section two and with 2T physics in section three, we may expect
that in a momentum space with a non-trivial geometry, such as the de Sitter
space in Snyder's momentum space approach, a natural choice is $h(p)=\det 
\bar{g}_{\mu \nu }$, where $\bar{g}_{\mu \nu }$ is given by equation (2.14).
This leads to the idea that the wave-particle duality may require that
momentum dependent metric tensors be present in the most general momentum
space formulation of quantum mechanics in the presence of gravity, in the
same way as position dependent metric tensors are present in the most
general position space formulation of quantum mechanics in the presence of
gravity described in [20]. But we also need another central object to
complete the formal structure of quantum mechanics. We need the concept of a
momentum dependent vector field $A_{\alpha }(p)$ with a vanishing strength
tensor in momentum space.

As a consequence of (4.8) and of our second assumption, we have the spectral
decomposition of the identity operator in the momentum eigenbasis $\mid
p\rangle $ 
\begin{equation}
1=\int_{D(p)}d^{n}p\sqrt{h(p)}\mid p\rangle \langle p\mid  \tag{4.9}
\end{equation}
This leads to the momentum space wave functions $\psi (p)=\langle p\mid \psi
\rangle $ and $\langle \psi \mid p\rangle =\langle p\mid \psi \rangle ^{\ast
}=\psi ^{\ast }(p)$ of any state $\mid \psi $ $\rangle $ belonging to the
Heisenberg algebra representation space 
\begin{equation}
\mid \psi \rangle =\int_{D(p)}d^{n}p\sqrt{h(p)}\psi (p)\mid p\rangle 
\tag{4.10a}
\end{equation}
\begin{equation}
\langle \psi \mid =\int_{D(p)}d^{n}p\sqrt{h(p)}\psi ^{\ast }(p)\langle p\mid
\tag{4.10b}
\end{equation}
The inner product of two states $\mid \psi \rangle $ and $\mid \varphi
\rangle $ is given in terms of their momentum space wave functions $\psi (p)$
and $\varphi (p)$ as 
\begin{equation}
\langle \psi \mid \varphi \rangle =\int_{D(p)}d^{n}p\sqrt{h(p)}\psi ^{\ast
}(p)\varphi (p)  \tag{4.11}
\end{equation}

The most general wave function $\langle x\mid p\rangle $ is given by [20] 
\begin{equation}
\langle x\mid p\rangle =\frac{e^{i\varphi (x_{0},p)}}{(2\pi \hbar )^{\frac{n%
}{2}}}\frac{\Omega \lbrack P(x_{0}\rightarrow x)]}{g^{\frac{1}{4}}(x)h^{%
\frac{1}{4}}(p)}e^{\frac{i}{\hbar }(x-x_{0}).p}  \tag{4.12}
\end{equation}
$\varphi (x_{0},p)$ is a specific but otherwise arbitrary real function and $%
\Omega \lbrack P(x_{0}\rightarrow x)]$ is the path ordered U(1) holonomy
along the path $P(x_{0}\rightarrow x)$. Notice that $g(x)$ and $h(p)$ are
both necessary because they simultaneously appear in the most general wave
function (4.12). The wave function (4.12) generalizes in a transparent
manner the usual plane wave solutions of application to the trivial
representation of the Heisenberg algebra with $A_{\alpha }(x)=0$ and with
the choices $g(x)=1$ and $h(p)=1$.

Now we point out that the wave-particle duality can be made explicit in
quantum mechanics if we introduce the equations 
\begin{equation}
\langle p\mid \hat{p}_{\alpha }\mid \psi \rangle =p_{\alpha }\langle p\mid
\psi \rangle =p_{\alpha }\psi (p)  \tag{4.13a}
\end{equation}
\begin{equation}
\langle p\mid \hat{x}_{\alpha }\mid \psi \rangle =\frac{i\hbar }{h^{1/4}(p)}[%
\frac{\partial }{\partial p^{\alpha }}+\frac{i}{\hbar }A_{\alpha }(p)%
]h^{1/4}(p)\psi (p)  \tag{4.13b}
\end{equation}
which are the momentum space correspondents of equations (4.6). The vector
field $A_{\alpha }(p)$ has a vanishing strength tensor in momentum space, 
\begin{equation}
\bar{F}_{\alpha \beta }=\frac{\partial A_{\beta }}{\partial p^{\alpha }}-%
\frac{\partial A_{\alpha }}{\partial p^{\beta }}=0  \tag{4.14}
\end{equation}
and is related to arbitrary local phase transformations of the momentum
eigenvectors 
\begin{equation}
\mid p^{\prime }\rangle =e^{\frac{i}{\hbar }\gamma (p)}\mid p\rangle 
\tag{4.15a}
\end{equation}
when 
\begin{equation}
A_{\alpha }^{\prime }(p)=A_{\alpha }(p)+\frac{\partial \gamma (p)}{\partial p%
}  \tag{4.15b}
\end{equation}
where $\gamma (p)$ is an arbitrary scalar function. Equations (4.15) are the
momentum space correspondents of equations (4.7). Now we need one evidence
that the complementation of the basic equations of quantum mechanics we
proposed in this section is really necessary. The best place to search for
this evidence is in 2T physics. First, because 2T physics is based on the
local indistinguishability of position and momentum. And second, because it
is now clear that 1T physics is embedded in 2T physics [18].

\section{2T Physics and Quantum Mechanics}

In section three we saw that position dependent and momentum dependent
metric tensors naturally appear in 2T physics. To show that position
dependent and momentum dependent vector fields also have a natural existence
and a unified description in 2T physics, we first modify the 2T Hamiltonian
(3.5) according to the usual minimal coupling prescription to position
dependent vector fields, $P_{M}\rightarrow P_{M}-A_{M}(X)$. Action (3.4b)
then becomes 
\begin{equation}
S=\int d\tau \{\dot{X}.P-[\frac{1}{2}\lambda _{1}(P-A)^{2}+\lambda
_{2}X.(P-A)+\frac{1}{2}\lambda _{3}X^{2}]\}  \tag{5.1}
\end{equation}
The equations of motion for the Lagrange multipliers now give the
constraints 
\begin{equation}
\phi _{1}=\frac{1}{2}(P-A)^{2}\approx 0  \tag{5.2}
\end{equation}
\begin{equation}
\phi _{2}=X.(P-A)\approx 0  \tag{5.3}
\end{equation}
\begin{equation}
\phi _{3}=\frac{1}{2}X^{2}\approx 0  \tag{5.4}
\end{equation}
The Poisson brackets between the canonical variables and the vector field $%
A_{M}(X)$ are 
\begin{equation}
\{X_{M},A_{N}\}=0  \tag{5.5a}
\end{equation}
\begin{equation}
\{P_{M},A_{N}\}=-\frac{\partial A_{N}}{\partial X^{M}}  \tag{5.5b}
\end{equation}
\begin{equation}
\{A_{M},A_{N}\}=0  \tag{5.5c}
\end{equation}
Computing the algebra of constraints (5.2)-(5.4) using the Poisson brackets
(3.9) and (5.5) we find the expressions 
\begin{equation}
\{\phi _{1},\phi _{1}\}=(P^{M}-A^{M})F_{MN}(P^{N}-A^{N})  \tag{5.6a}
\end{equation}
\begin{equation*}
\{\phi _{1},\phi _{2}\}=-2\phi _{1}+(P^{M}-A^{M})\frac{\partial }{\partial
X^{M}}(X.A)-(P-A).A
\end{equation*}
\begin{equation}
-X^{M}\frac{\partial }{\partial X^{M}}[(P-A).A]-X^{M}\frac{\partial }{%
\partial X^{M}}(\frac{1}{2}A^{2})  \tag{5.6b}
\end{equation}
\begin{equation}
\{\phi _{2},\phi _{2}\}=X^{M}F_{MN}X^{N}  \tag{5.6c}
\end{equation}
\begin{equation}
\{\phi _{1},\phi _{3}\}=-\phi _{2}  \tag{5.6d}
\end{equation}
\begin{equation}
\{\phi _{2},\phi _{3}\}=-2\phi _{3}  \tag{5.6e}
\end{equation}
\begin{equation}
\{\phi _{3},\phi _{3}\}=0  \tag{5.6f}
\end{equation}
For the purposes of this paper, we see from the above equations that
constraints (5.2)-(5.4) can be turned into first class constraints if the
vector field $A_{M}(X)$ satisfies the subsidiary conditions 
\begin{equation}
F_{MN}=\frac{\partial A_{N}}{\partial X^{M}}-\frac{\partial A_{M}}{\partial
X^{N}}=0  \tag{5.7}
\end{equation}
\begin{equation}
X.A=0  \tag{5.8a}
\end{equation}
\begin{equation}
(P-A).A=0  \tag{5.8b}
\end{equation}
\begin{equation}
\frac{1}{2}A^{2}=0  \tag{5.8c}
\end{equation}
Condition (5.7) implies that the vector field $A_{M}(X)$ defines a section
of a flat U(1) bundle over the $d+2$ dimensional position space. In the case
of vector fields for which $F_{MN}\neq 0$, condition (5.7) must be replaced
by the subsidiary conditions 
\begin{equation}
X^{M}F_{MN}=0  \tag{5.9a}
\end{equation}
\begin{equation}
(P^{M}-A^{M})F_{MN}=0  \tag{5.9b}
\end{equation}
Condition (5.9a) appeared first in [5] but if we use it alone we are not
taking into account the indistinguishability of $X_{M}$ and $P_{M}-A_{M}$ in
the presence of the vector field $A_{M}(X)$ for which $F_{MN}\neq 0$. As we
see from brackets (5.6), both conditions (5.9) are necessary in this case.

Conditions (5.8a)-(5.8c) imply that constraints (5.2)-(5.4) do not form an
irreducible [23] set of constraints for 2T physics with a vector field $%
A_{M}(X)$. Combining conditions (5.8a)-(5.8c) with constraints (5.2)-(5.4)
we obtain the irreducible set of constraints [19] 
\begin{equation}
\phi _{1}=\frac{1}{2}P^{2}\approx 0\text{ \ \ \ \ }\phi _{2}=X.P\approx 0%
\text{ \ \ }\phi _{3}=\frac{1}{2}X^{2}\approx 0  \tag{5.10a}
\end{equation}
\begin{equation}
\phi _{4}=X.A\approx 0\text{ \ \ \ }\phi _{5}=P.A\approx 0\text{ \ \ \ }\phi
_{6}=\frac{1}{2}A^{2}\approx 0  \tag{5.10b}
\end{equation}
The reappearance of constraints (5.10a) explains why the vector field $%
A_{M}(X)$ of vanishing strength tensor must be present in 2T physics. As we
saw in section three, in order to have a regular constraint surface
associated to constraints (5.10a), the origin of position space (viewed as
part of phase space) must be removed and this creates a non-trivial topology.

It can be verified that constraints (5.10) are all first class. The 2T
action in the presence of the vector field $A_{M}(X)$ can then be written as 
\begin{equation*}
S=\int d\tau \lbrack \dot{X}.P-(\frac{1}{2}\lambda _{1}P^{2}+\lambda _{2}X.P+%
\frac{1}{2}\lambda _{3}X^{2}
\end{equation*}
\begin{equation}
+\lambda _{4}X.A+\lambda _{5}P.A+\frac{1}{2}\lambda _{6}A^{2})]  \tag{5.11}
\end{equation}
where the 2T Hamiltonian is 
\begin{equation*}
H=\frac{1}{2}\lambda _{1}P^{2}+\lambda _{2}X.P+\frac{1}{2}\lambda _{3}X^{2}
\end{equation*}
\begin{equation}
+\lambda _{4}X.A+\lambda _{5}P.A+\frac{1}{2}\lambda _{6}A^{2}  \tag{5.12}
\end{equation}
It is important to mention here that, to arrive at action (511), no use was
made of conditions (5.7) and (5.9). Only conditions (5.8) were used. Action
(5.11) therefore gives a unified description of 2T physics with all kinds of
position dependent vector fields. Those for which $F_{MN}\neq 0$ or those
for which $F_{MN}=0$.

Action (5.11) is invariant under the Lorentz $SO(d,2)$ transformation with
generator $L_{MN}=X_{M}P_{N}-X_{N}P_{M}$ 
\begin{equation}
\delta X_{M}=\frac{1}{2}\omega _{RS}\{L_{RS},X_{M}\}=\omega _{MR}X_{R} 
\tag{5.13a}
\end{equation}
\begin{equation}
\delta P_{M}=\frac{1}{2}\omega _{RS}\{L_{RS},P_{M}\}=\omega _{MR}P_{R} 
\tag{5.13b}
\end{equation}
\begin{equation}
\delta A_{M}=\frac{\partial A_{M}}{\partial X_{R}}\delta X_{R}  \tag{5.13c}
\end{equation}
\begin{equation}
\delta \lambda _{\varrho }=0,\text{ \ \ \ \ \ }\varrho =1,2,...,6 
\tag{5.13d}
\end{equation}
under which $\delta S=0$. It can be checked that $L_{MN}$ has weakly
vanishing Poisson brackets with the first class constraints (5.10), being
therefore also gauge invariant in the presence of the vector field $A_{M}(X)$%
.

Action (5.11) also has the local infinitesimal invariance generated by the
first class constraints (5.10) 
\begin{equation}
\delta X_{M}=\epsilon _{\varrho }(\tau )\{X_{M},\phi _{\varrho }\}=\epsilon
_{1}P_{M}+\epsilon _{2}X_{M}+\epsilon _{5}A_{M}  \tag{5.14a}
\end{equation}
\begin{equation*}
\delta P_{M}=\epsilon _{\varrho }(\tau )\{P_{M},\phi _{\varrho }\}=-\epsilon
_{2}P_{M}-\epsilon _{3}X_{M}-\epsilon _{4}A_{M}
\end{equation*}
\begin{equation}
-\epsilon _{4}X^{N}\frac{\partial A_{N}}{\partial X^{M}}-\epsilon _{5}P^{N}%
\frac{\partial A_{N}}{\partial X^{M}}-\epsilon _{6}A^{N}\frac{\partial A_{N}%
}{\partial X^{M}}  \tag{5.14b}
\end{equation}
\begin{equation}
\delta A_{M}=\frac{\partial A_{M}}{\partial X^{N}}\delta X_{N}  \tag{5.14c}
\end{equation}
\begin{equation}
\delta \lambda _{1}=\dot{\epsilon}_{1}+2\epsilon _{2}\lambda _{1}-2\epsilon
_{1}\lambda _{2}  \tag{5.14d}
\end{equation}
\begin{equation}
\delta \lambda _{2}=\dot{\epsilon}_{2}+\epsilon _{3}\lambda _{1}-\epsilon
_{1}\lambda _{3}  \tag{5.14e}
\end{equation}
\begin{equation}
\delta \lambda _{3}=\dot{\epsilon}_{3}+2\epsilon _{3}\lambda _{2}-2\epsilon
_{2}\lambda _{3}  \tag{5.14f}
\end{equation}
\begin{equation}
\delta \lambda _{4}=\dot{\epsilon}_{4}+\epsilon _{3}\lambda _{5}-\epsilon
_{5}\lambda _{3}  \tag{5.14g}
\end{equation}
\begin{equation}
\delta \lambda _{5}=\dot{\epsilon}_{5}+\epsilon _{2}\lambda _{5}-\epsilon
_{5}\lambda _{2}  \tag{5.14h}
\end{equation}
\begin{equation}
\delta \lambda _{6}=\dot{\epsilon}_{6}  \tag{5.14i}
\end{equation}
under which 
\begin{equation}
\delta S=\int_{\tau _{i}}^{\tau _{f}}d\tau \frac{d}{d\tau }(\epsilon
_{\varrho }\phi _{\varrho })  \tag{5.15}
\end{equation}
Now the conserved charge, or the generator of the local transformations
(5.14), depending on wether the equations of motion are satisfied or not, is
the quantity $Q=\epsilon _{\varrho }\phi _{\varrho }$ with $\varrho
=1,2,...,6.$ This generalizes the local $Sp(2,R)$ invariance (3.11) of 2T
physics to the case when a vector field $A_{M}(X)$ is present.

Hamiltonian (5.12) is invariant under the finite local scale transformations 
\begin{equation}
X_{M}\rightarrow \exp \{\beta (\tau )\}X_{M}  \tag{5.16a}
\end{equation}
\begin{equation}
P_{M}\rightarrow \exp \{-\beta (\tau )\}P_{M}  \tag{5.16b}
\end{equation}
\begin{equation}
A_{M}\rightarrow \exp \{-\beta (\tau )\}A_{M}  \tag{5.16c}
\end{equation}
\begin{equation}
\lambda _{1}\rightarrow \exp \{2\beta (\tau )\}\lambda _{1}  \tag{5.16d}
\end{equation}
\begin{equation}
\lambda _{2}\rightarrow \lambda _{2}  \tag{5.16e}
\end{equation}
\begin{equation}
\lambda _{3}\rightarrow \exp \{-2\beta (\tau )\}\lambda _{3}  \tag{5.16f}
\end{equation}
\begin{equation}
\lambda _{4}\rightarrow \lambda _{4}  \tag{5.16g}
\end{equation}
\begin{equation}
\lambda _{5}\rightarrow \exp \{2\beta (\tau )\}\lambda _{5}  \tag{5.16h}
\end{equation}
\begin{equation}
\lambda _{6}\rightarrow \exp \{2\beta (\tau )\}\lambda _{6}  \tag{5.16i}
\end{equation}
We can use the local scale invariance (5.16) to again select the gauge where 
$\beta =\frac{1}{2}P^{2}$. In this gauge we have the bracket relations 
\begin{equation}
\{P_{M},P_{N}\}=0  \tag{5.17a}
\end{equation}
\begin{equation}
\{X_{M},P_{N}\}=\bar{G}_{MN}  \tag{5.17b}
\end{equation}
\begin{equation}
\{X_{M},X_{N}\}=-L_{MN}  \tag{5.17c}
\end{equation}
\begin{equation}
\{X_{M},A_{N}\}=-P_{M}A_{N}-X_{M}P^{S}\frac{\partial A_{N}}{\partial X^{S}} 
\tag{5.17d}
\end{equation}
\begin{equation}
\{P_{M},A_{N}\}=-\frac{\partial A_{N}}{\partial X^{M}}+P_{M}P^{S}\frac{%
\partial A_{N}}{\partial X^{S}}  \tag{5.17e}
\end{equation}
\begin{equation}
\{A_{M},A_{N}\}=A_{M}P^{S}\frac{\partial A_{N}}{\partial X^{S}}-A_{N}P^{S}%
\frac{\partial A_{M}}{\partial X^{S}}  \tag{5.17f}
\end{equation}
where $\bar{G}_{MN}$ is given by (3.17). The bracket relations (5.17) must
be used in the place of the Poisson brackets (3.9) and (5.5) when performing
classical Hamiltonian dynamics in the momentum space background $\bar{G}%
_{MN} $ with a vector field $A_{M}(X)$. In the transition to the quantized
theory using the correspondence principle, brackets (5.17) will give the
fundamental commutators for a formulation of quantum mechanics based on the
scale invariant Hamiltonian (5.12) where the geometry depends on the momenta
while the vector field depends on the positions. This new mixed formulation
must be an equally valid one because, as we will see below, the brackets
(5.17) preserve the form of the classical Hamiltonian equations of motion.

To reach a representation where both the geometry and the vector field are
position dependent, we use the local scale invariance (5.16) to select a
gauge where $\beta =\frac{1}{2}X^{2}$. In this gauge we have the bracket
relations 
\begin{equation}
\{P_{M},P_{N}\}=L_{MN}  \tag{5.18a}
\end{equation}
\begin{equation}
\{X_{M},P_{N}\}=G_{MN}  \tag{5.18b}
\end{equation}
\begin{equation}
\{X_{M},X_{N}\}=0  \tag{5.18c}
\end{equation}
\begin{equation}
\{X_{M},A_{N}\}=0  \tag{5.18d}
\end{equation}
\begin{equation}
\{P_{M},A_{N}\}=-\frac{\partial A_{N}}{\partial X^{M}}+X_{M}A_{N} 
\tag{5.18e}
\end{equation}
\begin{equation}
\{A_{M},A_{N}\}=0  \tag{5.18f}
\end{equation}
where $G_{MN}$ is given by (3.18). In the transition to the quantized theory
brackets (5.18) are turned into the fundamental commutators for a position
space formulation of quantum mechanics based on the same scale invariant
Hamiltonian (5.12).

In terms of the Poisson brackets (3.9) and (5.5), the classical equations of
motion in the presence of the vector field $A_{M}(X)$ are 
\begin{equation}
\dot{X}_{M}=\{X_{M},H\}=\lambda _{1}P_{M}+\lambda _{2}X_{M}+\lambda _{5}A_{M}
\tag{5.19a}
\end{equation}
\begin{equation*}
\dot{P}_{M}=\{P_{M},H\}=-\lambda _{2}P_{M}-\lambda _{3}X_{M}-\lambda
_{4}A_{M}
\end{equation*}
\begin{equation}
-\lambda _{4}X^{N}\frac{\partial A_{N}}{\partial X^{M}}-\lambda _{5}P^{N}%
\frac{\partial A_{N}}{\partial X^{M}}-\lambda _{6}A^{N}\frac{\partial A_{N}}{%
\partial X^{M}}  \tag{5.19b}
\end{equation}
\begin{equation}
\dot{A}_{M}=\{A_{M},H\}=\lambda _{1}P^{N}\frac{\partial A_{M}}{\partial X^{N}%
}+\lambda _{2}X^{N}\frac{\partial A_{M}}{\partial X^{N}}+\lambda _{5}A^{N}%
\frac{\partial A_{M}}{\partial X^{N}}  \tag{5.19c}
\end{equation}
where $H$ is given by (5.12). Since the Hamiltonian (5.12) is scale
invariant, after dropping terms quadratic in the constraints (5.10), it has
the same expression in the backgrounds $\bar{G}_{MN}$ and $G_{MN}$. However,
each background requires its own bracket structure. As can be verified,
after dropping terms linear in the constraints (5.10), the equations of
motion (5.19) remain invariant if we change to the momentum space background 
$\bar{G}_{MN}$ and use the brackets (5.17). After dropping terms linear in
the constraints (5.10), the equations of motion (5.19) also remain invariant
if we change to the position space background $G_{MN}$ while at the same
time changing to the bracket relations (5.18). Up to now we may say that
using the local scale invariance (5.16) we have uncovered three formulations
of quantum mechanics in three different spaces and with $A_{M}=A_{M}(X)$.
These three formulations have the same classical limit described by the 2T
Hamiltonian (5.12).

Now we use the local indistinguishability of position and momentum in 2T
physics and modify the 2T Hamiltonian (3.5) according to the new rule $%
X_{M}\rightarrow X_{M}-A_{M}(P).$ Action (3.4b) then becomes 
\begin{equation}
S=\int d\tau \lbrack \dot{X}.P-(\frac{1}{2}\lambda _{1}P^{2}+\lambda
_{2}(X-A).P+\frac{1}{2}\lambda _{3}(X-A)^{2}]  \tag{5.20}
\end{equation}
The equations of motion for the Lagrange multipliers now give the
constraints 
\begin{equation}
\phi _{1}=\frac{1}{2}P^{2}\approx 0  \tag{5.21}
\end{equation}
\begin{equation}
\phi _{2}=(X-A).P\approx 0  \tag{5.22}
\end{equation}
\begin{equation}
\phi _{3}=\frac{1}{2}(X-A)^{2}\approx 0  \tag{5.23}
\end{equation}
The Poisson brackets between the canonical variables and the vector field $%
A_{M}(P)$ are 
\begin{equation}
\{X_{M},A_{N}\}=\frac{\partial A_{N}}{\partial P^{M}}  \tag{5.24a}
\end{equation}
\begin{equation}
\{P_{M},A_{N}\}=0  \tag{5.24b}
\end{equation}
\begin{equation}
\{A_{M},A_{N}\}=0  \tag{5.24c}
\end{equation}
Computing the algebra of constraints (5.21)-(5.23) using the Poisson
brackets (3.9) and (5.24), we find the expressions 
\begin{equation}
\{\phi _{1},\phi _{1}\}=0  \tag{5.25a}
\end{equation}
\begin{equation}
\{\phi _{1},\phi _{2}\}=-2\phi _{1}  \tag{5.25b}
\end{equation}
\begin{equation}
\{\phi _{1},\phi _{3}\}=-\phi _{2}  \tag{5.25c}
\end{equation}
\begin{equation}
\{\phi _{2},\phi _{2}\}=-P^{M}\bar{F}_{MN}P^{N}  \tag{5.25d}
\end{equation}
\begin{equation*}
\{\phi _{2},\phi _{3}\}=-2\phi _{3}-P_{M}\frac{\partial }{\partial P_{M}}[%
(X-A).A]-P_{M}\frac{\partial }{\partial P_{M}}(\frac{1}{2}A^{2})
\end{equation*}
\begin{equation}
+(X_{M}-A_{M})\frac{\partial }{\partial P_{M}}(P.A)-(X-A).A  \tag{5.25e}
\end{equation}
\begin{equation}
\{\phi _{3},\phi _{3}\}=-(X^{M}-A^{M})\bar{F}_{MN}(X^{N}-A^{N})  \tag{5.25f}
\end{equation}
For the case in which we are interested in this paper, we see that
constraints (5.21)-(5.23) can be turned into first class constraints if we
impose the subsidiary conditions on the vector field $A_{M}(P)$ 
\begin{equation}
\tilde{F}_{MN}=\frac{\partial A_{N}}{\partial P^{M}}-\frac{\partial A_{M}}{%
\partial P^{N}}=0  \tag{5.26}
\end{equation}
\begin{equation}
(X-A).A=0  \tag{5.27a}
\end{equation}
\begin{equation}
P.A=0  \tag{5.27b}
\end{equation}
\begin{equation}
\frac{1}{2}A^{2}=0  \tag{5.27c}
\end{equation}
Condition (5.26) implies that the vector field $A_{M}(P)$ defines a section
of a flat U(1) bundle over the momentum space. For vector fields $A_{M}(P)$
for which $\bar{F}_{MN}\neq 0$ the subsidiary conditions can be obtained
from brackets (5.25f) and (5.25d) and are 
\begin{equation}
(X^{M}-A^{M})\bar{F}_{MN}=0  \tag{5.28a}
\end{equation}
\begin{equation}
P^{M}\bar{F}_{MN}=0  \tag{5.28b}
\end{equation}
Compare conditions (5.28) with conditions (5.9) that were obtained in the
case of a vector field $A_{M}(X)$ for which $F_{MN}\neq 0$. There is a clear
dual relation between (5.9) and (5.28).

Conditions (5.27a)-(5.27c) imply that constraints (5.21)-(5.23) do not form
an irreducible set of constraints for 2T physics with a vector field $%
A_{M}(P)$. Combining conditions (5.27a)-(5.27c) with constraints
(5.21)-(5.23) we arrive at the same set of irreducible first class
constraints (5.10). Action (5.11) therefore gives a unified general
description of 2T physics with position dependent or momentum dependent
vector fields because again no use was made of conditions (5.26) and (5.28)
to arrive at action (5.11).

To each of the symmetries (5.13), (5.14) and (5.16) of action (5.11) with $%
A_{M}=A_{M}(X)$ there is a corresponding symmetry with $A_{M}=A_{M}(P).$ The
global $SO(d,2)$ invariance with generator $L_{MN}=X_{M}P_{N}-X_{N}P_{M}$ is
given by the transformation equations 
\begin{equation}
\delta X_{M}=\frac{1}{2}\omega _{RS}\{L_{RS},X_{M}\}=\omega _{MR}X_{R} 
\tag{5.29a}
\end{equation}
\begin{equation}
\delta P_{M}=\frac{1}{2}\omega _{RS}\{L_{RS},P_{M}\}=\omega _{MR}P_{R} 
\tag{5.29b}
\end{equation}
\begin{equation}
\delta A_{M}=\frac{\partial A_{M}}{\partial P_{N}}\delta P_{N}  \tag{5.29c}
\end{equation}
\begin{equation}
\delta \lambda _{\varrho }=0\text{ \ \ \ \ \ }\varrho =1,2,.....,6 
\tag{5.29d}
\end{equation}
under which $\delta S=0$. It can be verified that $L_{MN}$ has weakly
vanishing Poisson brackets with constraints (5.10) when $A_{M}=A_{M}(P)$,
being therefore also gauge invariant in this case.

The first class constraints (5.10) generate the local infinitesimal
transformations 
\begin{equation*}
\delta X_{M}=\epsilon _{\varrho }(\tau )\{X_{M},\phi _{\varrho }\}=\epsilon
_{1}P_{M}+\epsilon _{2}X_{M}+\epsilon _{4}X^{S}\frac{\partial A_{S}}{%
\partial P^{M}}
\end{equation*}
\begin{equation}
+\epsilon _{5}A_{M}+\epsilon _{5}P^{S}\frac{\partial A_{S}}{\partial P^{M}}%
+\epsilon _{6}A^{S}\frac{\partial A_{S}}{\partial P^{M}}  \tag{5.30a}
\end{equation}
\begin{equation}
\delta P_{M}=\epsilon _{\varrho }(\tau )\{P_{M},\phi _{\varrho }\}=-\epsilon
_{2}P_{M}-\epsilon _{3}X_{M}-\epsilon _{4}A_{M}  \tag{5.30b}
\end{equation}
\begin{equation}
\delta A_{M}=\frac{\partial A_{M}}{\partial P_{N}}\delta P_{N}  \tag{5.30c}
\end{equation}
\begin{equation}
\delta \lambda _{1}=\dot{\epsilon}_{1}+2\epsilon _{2}\lambda _{1}-2\epsilon
_{1}\lambda _{2}  \tag{5.30d}
\end{equation}
\begin{equation}
\delta \lambda _{2}=\dot{\epsilon}_{2}+\epsilon _{3}\lambda _{1}-\epsilon
_{1}\lambda _{3}  \tag{5.30e}
\end{equation}
\begin{equation}
\delta \lambda _{3}=\dot{\epsilon}_{3}+2\epsilon _{3}\lambda _{2}-2\epsilon
_{2}\lambda _{3}  \tag{5.30f}
\end{equation}
\begin{equation}
\delta \lambda _{4}=\dot{\epsilon}_{4}+\epsilon _{4}\lambda _{2}-\epsilon
_{2}\lambda _{4}  \tag{5.30g}
\end{equation}
\begin{equation}
\delta \lambda _{5}=\dot{\epsilon}_{5}+\epsilon _{4}\lambda _{1}-\epsilon
_{1}\lambda _{4}  \tag{5.30h}
\end{equation}
\begin{equation}
\delta \lambda _{6}=\dot{\epsilon}_{6}  \tag{5.30i}
\end{equation}
under which 
\begin{equation}
\delta S=\int_{\tau _{i}}^{\tau _{f}}d\tau \frac{d}{d\tau }(\epsilon
_{\varrho }\phi _{\varrho })  \tag{5.31}
\end{equation}
The conserved Hamiltonian Noether charge $Q$, or the generator of the local
transformations (5.30), depending on wether the equations of motion are
satisfied or not, is again a combination of the first class constraints
(5.10), with the exception that now $A_{M}=A_{M}(P)$.

In the case when $A_{M}=A_{M}(P)$, Hamiltonian (5.12) is invariant under the
finite local scale transformations 
\begin{equation}
X_{M}\rightarrow \exp \{\beta (\tau )\}X_{M}  \tag{5.32a}
\end{equation}
\begin{equation}
P_{M}\rightarrow \exp \{-\beta (\tau )\}P_{M}  \tag{5.32b}
\end{equation}
\begin{equation}
A_{M}\rightarrow \exp \{\beta (\tau )\}A_{M}  \tag{5.32c}
\end{equation}
\begin{equation}
\lambda _{1}\rightarrow \exp \{2\beta (\tau )\}\lambda _{1}  \tag{5.32d}
\end{equation}
\begin{equation}
\lambda _{2}\rightarrow \lambda _{2}  \tag{5.32e}
\end{equation}
\begin{equation}
\lambda _{3}\rightarrow \exp \{-2\beta (\tau )\}\lambda _{3}  \tag{5.32f}
\end{equation}
\begin{equation}
\lambda _{4}\rightarrow \exp \{-2\beta (\tau )\}\lambda _{4}  \tag{5.32g}
\end{equation}
\begin{equation}
\lambda _{5}\rightarrow \lambda _{5}  \tag{5.32h}
\end{equation}
\begin{equation}
\lambda _{6}\rightarrow \exp \{-2\beta (\tau )\}\lambda _{6}  \tag{5.32i}
\end{equation}
Again we can use the local scale invariance (5.32) to select a gauge where $%
\beta =\frac{1}{2}P^{2}$. After manipulations similar to the case when $%
A_{M}=A_{M}(X)$, we arrive at the bracket relations 
\begin{equation}
\{P_{M},P_{N}\}=0  \tag{5.33a}
\end{equation}
\begin{equation}
\{X_{M},P_{N}\}=\bar{G}_{MN}  \tag{5.33b}
\end{equation}
\begin{equation}
\{X_{M},X_{N}\}=-L_{MN}  \tag{5.33c}
\end{equation}
\begin{equation}
\{X_{M},A_{N}\}=\frac{\partial A_{N}}{\partial P^{M}}+P_{M}A_{N}  \tag{5.33d}
\end{equation}
\begin{equation}
\{P_{M},A_{N}\}=0  \tag{5.33e}
\end{equation}
\begin{equation}
\{A_{M},A_{N}\}=0  \tag{5.33f}
\end{equation}
In the transition to the quantized theory using the correspondence
principle, the brackets (5.33) will give the fundamental commutators for a
formulation of quantum mechanics in the presence of gravity based on the
scale invariant Hamiltonian (5.12) in a momentum space with a non-trivial
topology. In this formulation the metric tensor and the vector field are
both momentum dependent. If we assume that 1T physics is embedded [18] in 2T
physics, the existence of the $d+2$ dimensional brackets (5.33) can be used
to justify the complementation of the equations of quantum mechanics we
proposed in section four. The brackets (5.33) are the momentum space
correspondents of the position space brackets (5.18) we obtained above in
the case when $A_{M}=A_{M}(X)$

We can also use the local scale invariance (5.32) to select a gauge where $%
\beta =\frac{1}{2}X^{2}$. In this case we arrive at the brackets 
\begin{equation}
\{P_{M},P_{N}\}=L_{MN}  \tag{5.34a}
\end{equation}
\begin{equation}
\{X_{M},P_{N}\}=G_{MN}  \tag{5.34b}
\end{equation}
\begin{equation}
\{X_{M},X_{N}\}=0  \tag{5.34c}
\end{equation}
\begin{equation}
\{X_{M},A_{N}\}=\frac{\partial A_{N}}{\partial P^{M}}+X_{M}X^{S}\frac{%
\partial A_{N}}{\partial P^{S}}  \tag{5.34d}
\end{equation}
\begin{equation}
\{P_{M},A_{N}\}=-X_{M}A_{N}-P_{M}X^{S}\frac{\partial A_{N}}{\partial P^{S}} 
\tag{5.34e}
\end{equation}
. 
\begin{equation}
\{A_{M},A_{N}\}=A_{M}X^{S}\frac{\partial A_{N}}{\partial P^{S}}-A_{N}X^{S}%
\frac{\partial A_{M}}{\partial P^{S}}  \tag{5.34f}
\end{equation}
In the transition to the quantized theory, brackets (5.34) will give the
fundamental commutators for another formulation of quantum mechanics based
on the same scale invariant Hamiltonian (5.12). In this formulation the
geometry depends on the position while the vector field is momentum
dependent. This formulation is dual to the formulation that can be obtained
from brackets (5.17).

In the case of a momentum dependent vector field $A_{M}(P)$, Hamiltonian
(5.12) generates the equations of motion 
\begin{equation*}
\dot{X}_{M}=\{X_{M},H\}=\lambda _{1}P_{M}+\lambda _{2}X_{M}+\lambda _{5}A_{M}
\end{equation*}
\begin{equation}
+\lambda _{4}X^{N}\frac{\partial A_{N}}{\partial P^{M}}+\lambda _{5}P^{N}%
\frac{\partial A_{N}}{\partial P^{M}}+\lambda _{6}A^{N}\frac{\partial A_{N}}{%
\partial P^{M}}  \tag{5.35a}
\end{equation}
\begin{equation}
\dot{P}_{M}=\{P_{M},H\}=-\lambda _{2}P_{M}-\lambda _{3}X_{M}-\lambda
_{4}A_{M}  \tag{5.35b}
\end{equation}
\begin{equation}
\dot{A}_{M}=\{A_{M},H\}=-\lambda _{2}P^{N}\frac{\partial A_{M}}{\partial
P^{N}}-\lambda _{3}X^{N}\frac{\partial A_{M}}{\partial P^{N}}-\lambda
_{4}A^{N}\frac{\partial A_{M}}{\partial P^{N}}  \tag{5.35c}
\end{equation}
computed in terms of the Poisson brackets (3.9) and (5.24). The Hamiltonian
and the equations of motion remain invariant if we change to the background $%
\bar{G}_{MN}$ supplied with brackets (5.33). The Hamiltonian and the
equations of motion also remain invariant if we change to the background $%
G_{MN}$ supplied with brackets (5.34). Using the local scale invariance
(5.32) we have now uncovered three other formulations of quantum mechanics
in three different spaces and with $A_{M}=A_{M}(P)$. \ We must then conclude
that there are six possible formulations of quantum mechanics. These six
formulations have the same classical limit described by the 2T Hamiltonian
(5.12).

\section{ Concluding \ remarks}

Starting with the massless scalar relativistic particle, we provided
evidence for the necessity for the concept of a momentum dependent metric
tensor in quantum mechanics in the presence of gravity. Then we showed that
position and momentum dependent metric tensors have a natural existence in
2T physics as a consequence of a local scale invariance of the Hamiltonian.
Also as a consequence of this local scale invariance, we verified that the
classical Hamiltonian equations of motion for 2T physics are identical in
the backgrounds defined by these position dependent and momentum dependent
metric tensors. This demonstrates their equivalence at the classical level
in 2T physics.

Based on this equivalence of position and momentum dependent metric tensors
at the classical level in 2T physics, we wrote down the equations that makes
quantum mechanics in the presence of gravity in total agreement with the
wave-particle duality. In constructing these equations for the case of
momentum spaces with non-trivial topology, we had to introduce the concept
of a momentum dependent vector field. As a basis for this concept, this
paper verifies that the symmetries which are present in 2T physics when $%
A_{M}=A_{M}(X)$ are also present when $A_{M}=A_{M}(P)$.

\noindent

\noindent

\end{document}